\documentclass[fleqn,usenatbib]{mnras}

\usepackage{newtxtext,newtxmath}

\usepackage[T1]{fontenc}

\DeclareRobustCommand{\VAN}[3]{#2}
\let\VANthebibliography\thebibliography
\def\thebibliography{\DeclareRobustCommand{\VAN}[3]{##3}\VANthebibliography}

\usepackage{graphicx}	
\usepackage{amsmath}	

\title[Abell 370 radio halo]{MeerKAT reveals an ultra-steep spectrum radio halo in Abell 370}

\author[G. D. Lawrie et al.]{
Graham D. Lawrie$^{1,2}$ and
Roger P. Deane$^{1,3,4}$
\\
$^{1}$Wits Centre for Astrophysics, School of Physics, University of the Witwatersrand, 1 Jan Smuts Avenue, Johannesburg, 2000, South Africa \\
$^{2}$Department of Physics \& Astronomy, University of the Western Cape, Robert Sobukwe Rd, Bellville, 7535, South Africa\\
$^{3}$Inter-University Institute for Data Intensive Astronomy, Department of Astronomy, University of Cape Town, Cape Town, South Africa \\
$^{4}$Department of Physics, University of Pretoria, Private Bag X20, Pretoria 0028, South Africa 
}

\date{Accepted 2026 November 05. Received 2026 November 02; in original form 2026 January 22}

\pubyear{\the\year{}}

\begin{document}
\label{firstpage}
\pagerange{\pageref{firstpage}--\pageref{lastpage}}
\maketitle

\begin{abstract}
    Hubble Frontier Field cluster Abell 370 ($z=0.375$) has been the target of a suite of legacy multi-wavelength observational campaigns and is a prime candidate to host a diffuse radio halo, given its high mass ($\sim10^{15}\,\rm{M_\odot}$) and non-relaxed dynamical state. Previous detections of its radio halo were of low significance and imaging fidelity, leading to an uncertain characterisation of its nature. We use deep MeerKAT Ultra-High Frequency observations ($580-1015\,\rm{MHz}$) to robustly map the halo's spatial extent and spectral properties. We find a radio emission size of $\sim 1.1\,\rm{Mpc}$, significantly larger than previously reported. We measure a MeerKAT UHF in-band integrated spectral index of $\alpha=-1.54\pm0.05$. This  result is consistent with an ultra-steep spectrum radio halo ($\alpha\lesssim-1.5$). We find a strong, sub-linear correlation ($k=0.60\pm0.08$) between the radio and X-ray surface brightness. The radio halo's position on the $P_{1.4\,\rm{GHz}}-M_{500}$ scaling relation is underluminous, consistent with the known USSRH population. The combined evidence strongly supports that the diffuse central radio emission in Abell~370 is an Ultra-Steep Spectrum Radio Halo (USSRH), refining its previous classification and providing further support for the turbulent re-acceleration model.
\end{abstract}




\section{Introduction}
\label{intro}

Galaxy clusters are the largest gravitationally bound and virialised structures in the Universe, with halo masses of $10^{14}-10^{15}\,\rm{M_{\odot}}$. They are situated at the nodes of the cosmic web, where large-scale structure filaments intersect \citep{Aragon_2010}, and are assembled through the hierarchical merging of smaller galaxy groups and clusters \citep{Peebles_1980}. The primary component of these massive halos is dominated by dark matter, which makes up 70-80 per cent of the mass \citep{Blumenthal_1984}. The remaining mass is baryonic matter composed of the dominant intra-cluster medium (ICM), a hot ($10^7-10^8\,\rm{K}$), optically thin gas filling the cluster potential \citep{Mitchell_1976,Serlemitsos_1977}, and the negligible stellar mass in member galaxies.

Historically, the ICM has been traced by its thermal X-ray emission. However, modern radio telescopes have detected non-thermal synchrotron emission counterparts in a fraction of galaxy clusters, generated by the presence of relativistic particles in large-scale magnetic fields. This diffuse radio emission is broadly classified into four main categories: giant radio halos, mini-halos, radio relics, and radio phoenixes \citep[see][for a review]{Brunetti_2014,van_Weeren_2019}.

Giant radio halos (RHs) are spatially extended ($\gtrsim1\,\rm{Mpc}$), centrally located, and found predominantly in massive, dynamically disturbed (merging) systems \citep{Cassano_2013,Cuciti_2021b}. Typical RHs have spectral indices\footnote{Using the $S_{\nu}\propto\nu^{\alpha}$ convention} in the range of $\alpha=-1.1$ to $-1.4$ \citep{Giovannini_2009,Feretti_2012}. Their radio emission is known to spatially correlate with the ICM distribution, traced by thermal X-ray emission \citep{Govoni_2001a,Giacintucci_2005}. In the last two decades, a new population of ultra-steep spectrum radio halos (USSRHs) has been revealed, with steeper spectra of $\alpha\lesssim-1.5$, with their origin connected to less energetic mergers \citep{Brunetti_2008, Dallacasa_2009,Macario_2010,Bonafede_2012,Duchesne_2021,Cuciti_2021,Pasini_2022,Edler_2022,Venturi_2022,Rajpurohit_2023,magolego_2026}.

Radio relics are elongated, arc-shaped sources, often highly polarised and located in cluster periphery \citep{Ensslin_1998,Gasperin_2022,Pasini_2022}. Their origin is linked to particle acceleration from shock fronts, supported by their spatial coincidence with X-ray surface brightness and temperature discontinuities \citep{Akamatsu_2015, Eckert_2016}. 

A related relic classification, radio phoenixes, are smaller ($\lesssim400\,\rm{kpc}$) sources, thought to be "revived" from fossil lobes of past active galactic nucleus (AGN) activity. Most commonly, they have filamentary-like morphology, though this can vary \citep{Slee_2001}. Adiabatic compression from passing shocks is thought to re-energise the old plasma, resulting in steep power law spectra ($\alpha\lesssim-1.5$) sources closer to the cluster centre when compared to relics \citep{van_Weeren_2009,Feretti_2012}. 

Finally, mini-halos (MHs) are smaller-scale ($\lesssim500\,\rm{kpc}$), steep-spectrum, diffuse sources found at the centre of dynamically relaxed, cool-core clusters \citep{Giacintucci_2017,Kolokythas_2025}. Their origin is thought to be related to turbulent re-acceleration driven by gas "sloshing" in the cluster core, distinctly different from the high-energy turbulence associated with merging clusters \citep{Giacintucci_2014,Brunetti_2014,Giacintucci_2017}.

The origin of RHs is still debated, with two primary competing models: the turbulent re-acceleration model and the hadronic model. In the currently favoured turbulent re-acceleration model, a pre-existing population of electrons is accelerated by turbulent energy injected through merger events \citep{Ensslin_1998,Brunetti_2004}. In the hadronic model, relativistic electrons are produced as secondary particles from collisions between ICM and cosmic-ray protons \citep{Dennison_1980}. The existence of USSRHs supports the turbulent re-accleration model, as it is difficult to reproduce USSRHs in purely hadronic scenarios. While a combination of these mechanisms being responsible is a possibility \citep{Brunetti_2011}, next-generation radio telescopes will provide important constraints on the faint radio emission in relaxed galaxy clusters, helping determine the contributions of different mechanisms. It is important in this new era to build up large statistical samples of USSRHs with high fidelty mapping, broad spectral coverage, and well calibrated masses.

The galaxy cluster Abell 370 ($z=0.375$) is a massive unrelaxed system ($M_{\rm{500}}\sim10^{15}\,\rm{M_{\odot}}$) in a Hubble Frontier Field \citep{Abell_1989,Planck_2016,Lotz_2017}. It is renowned for being the first cluster discovered to gravitationally lens a background galaxy \citep{Hoag_1981,Soucail_1987,Paczynski_1987}, and has been extensively studied across the electromagnetic spectrum \citep{Broadhurst_2008,Richard_2010,Lagattuta_2017,Xie_2020, Kolokythas_2025}. Its dynamical state is a clear merger, with a bimodal mass distribution centred on two distinct brightest cluster galaxies (BCGs) \citep{Richard_2010}. Abell~370 is expected to host diffuse radio emission consistent with a typical RH, given its high mass and dynamically disturbed state. Diffuse radio emission in Abell 370 was first discovered by \citet{Xie_2020}, who detected a faint radio halo at $325\,\rm{MHz}$ and $1.5\,\rm{GHz}$ using Giant Metrewave Radio Telescope (GMRT) and Very Large Array (VLA) observations, and place an upper limit on the emission at $3\,\rm{GHz}$. They report an integrated spectral index of $\alpha^{1500}_{325}=-1.10\pm0.09$ and classify the source as a radio halo. At $1.28\,\rm{GHz}$ this faint emission was detected and classified as a candidate radio halo by \citet{Knowles_2022} as part of the MeerKAT Galaxy Cluster Legacy Survey \citep[MGCLS; see also][]{Kolokythas_2025}.

In this work, we begin by describing the data processing strategy in Sec.~\ref{data_reduc}. In Sec.~\ref{diff_radio}, we present the radio continuum images, which reveal a larger "largest linear size" (LLS) than previously reported and unknown radio extensions to the southeast and southwest not included in the literature. Sec.~\ref{ancillary} qualitatively explores the co-location of thermal X-ray emission and radio emission. In Sec.~\ref{spidx} we derive a steep spectral index consistent with an USSRH for Abell 370. We then quantitatively confirm the X-ray-radio co-location in Sec.~\ref{xray_discussion}, using a point-to-point region-based analysis. We investigate the parameter space in the power-mass scaling relation occupied by typical radio halos versus that of USSRHs in Sec.~\ref{scaling_relations}. Finally, we summarise our findings in Sec.~\ref{conclusions}.

\section{Observations and Data Reduction}
\label{data_reduc}

The data used in this work stems from MeerKAT open time data (project ID: SCI-20241101-GL-01), and is comprised of four observations of $\sim5\,\rm{hours}$ each. The observations were taken using the MeerKAT UHF-band ($580-1015\,\rm{MHz}$ frequency range and $544\,\rm{MHz}$ bandwidth) in the wideband fine 32K channelisation mode ($16.6\,\rm{kHz}$ channel width). All four observations use more than 61 of the 64 available antennas, have dump times of $\sim8\,\rm{sec}$, and observe the following sources: J0408-6545 (flux and bandpass calibrator), J0240-2309 (complex gains calibrator), and A370 (target galaxy cluster). The observations were made on the $21^{\rm{st}}$, $24^{\rm{th}}$, $27^{\rm{th}}$, and $28^{\rm{th}}$ of December 2024.

The MeerKAT UHF-band raw visibilities are reduced using the semi-automatic MeerKAT data reduction pipeline \textsc{oxkat} \citep{Heywood_2020}, which uses routines that are standard practice for MeerKAT data calibration, such as flagging, reference calibration (1GC), self-calibration (2GC), and direction-dependent calibration (3GC). \textsc{oxkat} uses tasks from \textsc{casa} \citep{McMullin_2007}, \textsc{cubical} \citep{Kenyon_2018}, and \textsc{killms} \citep{Tasse_2023}, while using \textsc{wsclean} \citep{Offringa_2014} and \textsc{ddfacet} \citep{Tasse_2023dd} for de-convolution and imaging at the 2GC and 3GC stage, respectively.

Together with the automated RFI excision software package, \textsc{tricolour} \citep{Hugo_2022}, we perform manual flagging of bad antennas, $(u,v)$ ranges, and occasionally calibrator scans, to remove radio frequency interference (RFI) and bad data. Cross calibration is done through \textsc{oxkat}, which uses standard \textsc{casa} tasks such as \texttt{gaincal} and \texttt{applycal} to correct for atmospheric phase shifts among other effects. Self-calibration is then done by using \textsc{cubical}, in conjunction with \textsc{wsclean}, to iteratively improve the cleaning mask, resulting in higher imaging fidelity. However, significant DDEs in the field, caused by bright off-axis sources, limit the improvements gained by self-calibration.

To mitigate the effects of these DDEs, we utilise the "peeling" functionality in the 3GC section of the \textsc{oxkat} pipeline. Peeling involves predicting visibilities for a problematic source, and for the full sky model, then subtracting the problematic visibilities from the full sky model with \textsc{cubical}. We sequentially peel six bright, off-centre sources using this method. The result is a significant deduction in DDEs in the field, and a significant improvement in both the imaging fidelity and dynamical range.

To accurately characterise diffuse radio emission, it is vital to correct for the flux contribution from compact sources hosted within the diffuse emission. To achieve this, we perform point source subtraction using \textsc{crystallball} \citep{Jozsa_2020}, \textsc{casa}'s \texttt{uvsub}, and \textsc{wsclean} in a similar manner to \citet{magolego_2026}. A high-resolution continuum image, produced with a \texttt{robust Briggs} weighting of -2, is used to identify and draw regions around point sources in and around the already visible, diffuse emission. We then use \textsc{crystallball} to generate visibilities for these identified point sources, which are subtracted from the calibrated visibilities using \texttt{uvsub}. This procedure is performed twice, as re-imaging reveals additional point sources due to reduced blending. The final, point-source free, visibilities are imaged with a \texttt{robust Briggs} weighting of 0.5 to produce a lower-resolution image which is more sensitive to the faint, diffuse, and extended radio emission. Despite the $uv$-plane subtraction, residual emission from bright point sources can remain in the image plane. To mitigate this we implement an additional image-plane masking technique where we apply spatial masks over regions affected by point-source emission. We then extrapolate the mean surface brightness of surrounding clean diffuse emission across the masked regions by interpolating the data using a 2D Gaussian kernel scaled to the synthesized beam size. This combined approach allows for a robust extraction of the radio surface brightness and total flux density across the halo emission.

\section{Results}

\subsection{Diffuse Radio Emission in Abell 370}
\label{diff_radio}

A crucial step in characterising radio halos in galaxy clusters is to model and subtract contaminating point sources (outlined in Sec.~\ref{data_reduc}). This isolates the diffuse radio emission, allowing for a study of its full extent and morphology. Here, we showcase the enhanced image fidelity achieved through meticulous subtraction, which enables the high-fidelity mapping of the full halo extent and features discussed here. 

\begin{figure*}
    \centering
    \includegraphics[width=\textwidth]{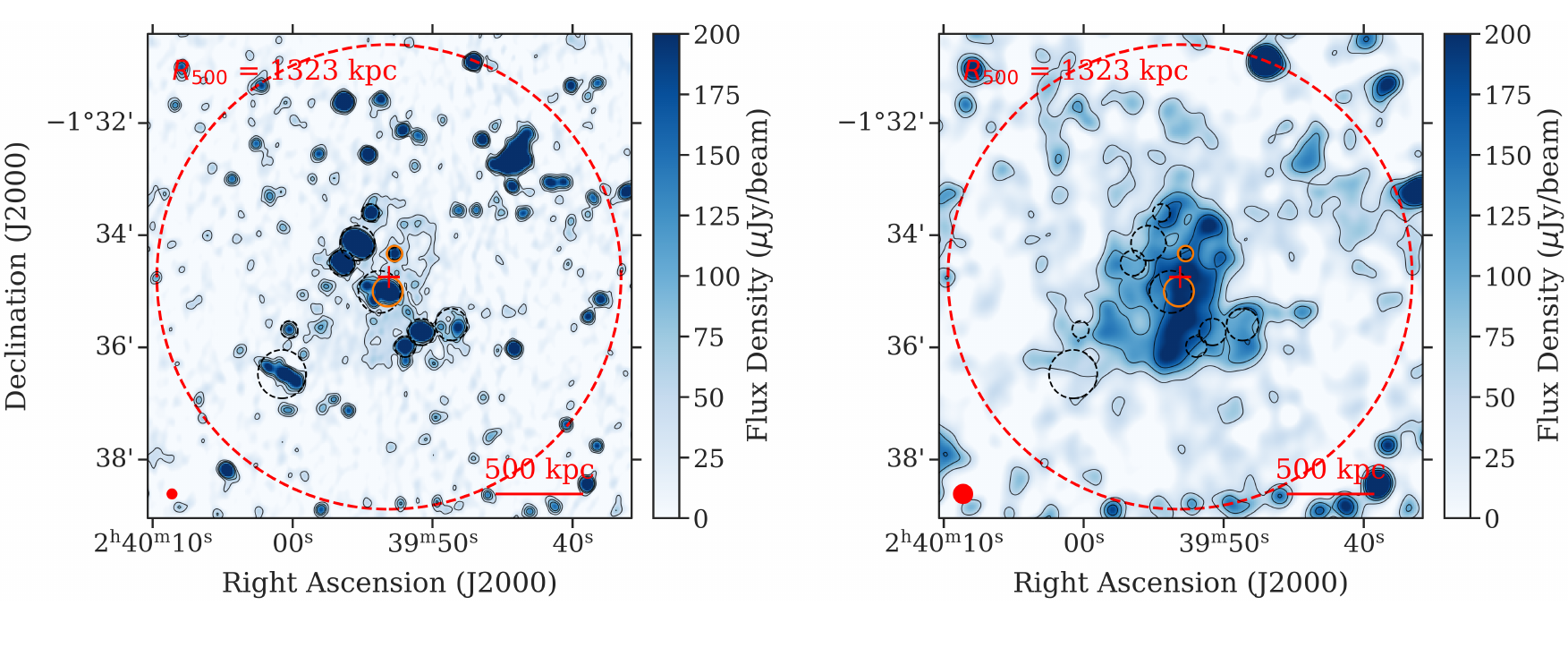}
    \caption{The full-resolution map before point source subtraction (left) and the lower-resolution map after point source subtraction (right) of the MeerKAT UHF ($816\,\rm{MHz}$ central frequency) radio emission detected in galaxy cluster Abell 370. The full-resolution map is imaged with a \texttt{robust} weighting of -0.5 and PSF dimensions of $10\times10\,\rm{arcsec}$, while the lower-resolution map is imaged with a \texttt{robust} weighting of 0 and tapered with a $14\,\rm{arcsec}$ Gaussian kernel, with a PSF of $20\times20\,\rm{arcsec}$, see \ref{data_reduc}. The $R_{500}$, PSF, and scale bar are shown in the top-left, bottom-left, and bottom-right of the images, respectively. Additionally, we encircle the two BCGs with solid orange lines, the $R_{500}$ with a dashed red line, 10 bright sources targeted for image-plane masking with dashed black circles, and indicate the ACT SZ detection centroid with a red cross \citep{Hilton_2021}. Radio contours are overlaid at [5,10,15]$\times\sigma_{\rm{rms}}=7\,\mu\rm{Jy}\,\rm{beam}^{-1}$ (left) and [5,10,15]$\times\sigma_{\rm{rms}}=10\,\mu\rm{Jy}\,\rm{beam}^{-1}$ (right). While visible in both images, the lower-resolution image clearly shows a diffuse radio halo with an LLS of $\sim1.1\,\rm{Mpc}$ at $5\sigma$ significance.}
    \label{fig:diffuse_emission}
\end{figure*}

In Fig.~\ref{fig:diffuse_emission} we show the MeerKAT UHF-band view of the radio halo in Abell 370. In the left panel, we show a higher-resolution (\texttt{robust Briggs}=-0.5) radio continuum map before point source subtraction with point spread function (PSF) dimensions of $10\times10\,\rm{arcsec}$, while the right panel shows the lower-resolution (\texttt{robust Briggs}=0) $14\,\rm{arcsec}$ Gaussian-tapered, point source subtracted image with a PSF of $21\times21\,\rm{arcsec}$, see Sec.\ref{data_reduc} for details. In both panels, we indicate the radius within which the mean total mass density is 500 times the critical density of the Universe ($R_{500}$), the PSF, a $500\,\rm{kpc}$ scale bar, and the Atacama Cosmology Telescope (ACT) SZ detection centroid \citep{Hilton_2021}. The physical $R_{500}$ radius is derived from the ACT DR5 cluster mass \citep{Hilton_2021}. Contours are overlaid at $[5,10,15]\times\sigma_{\rm{rms}}$ where $\sigma_{\rm{rms}}$ is determined using a nearby source-free region in the respective radio map, $\sigma_{\rm{rms}}=7$ and $10\,\mu\rm{Jy}\,\rm{beam}^{-1}$ for the full- and lower-resolution images, respectively. We encircle the BCGs to the north and south of the SZ centre in orange. Finally, pixels are colourised by radio brightness in units of $\mu\rm{Jy}\,\rm{beam}^{-1}$.

In the left panel of Fig.~\ref{fig:diffuse_emission}, we clearly see diffuse emission associated with Abell 370, traced by the $5\sigma$ contours. This is expected, given that Abell 370's merging state results in a dynamically unrelaxed cluster \citep{Buote_2001, Brunetti_2001}. As stated in previous studies, the diffuse emission in the left panel appears preferentially centred on the southernmost of the two merging clusters \citep{Xie_2020}. After point source subtraction and lower-resolution imaging, the radio halo becomes prominent at both the $10\sigma$ and the $15\sigma$ contours throughout the system, revealing significantly more extended and higher signal-to-noise-ratio (SNR) diffuse emission. The LLS, previously measured to be $\sim500-700\,\rm{kpc}$ \citep{Xie_2020} or $\sim950\,\rm{kpc}$ \citep{Kolokythas_2025}, is measured in both the full- and lower-resolution images to be $\sim1.1\,\rm{Mpc}$. To mitigate the impact of remaining low-level residuals from the $uv$-plane subtraction, we mask 10 point source regions (dashed black circles) as described previously. The point source subtraction appears successful, leaving prominent diffuse emission within the $R_{500}$ boundary.

Our point source subtraction and lower-resolution imaging have therefore revealed a larger LLS than previously measured. However, a multiwavelength approach, including complementary X-ray observations, is critical to fully understand the ongoing physical processes and to characterise this halo. 

\subsection{Multiwavelength Data}
\label{ancillary}

The multiwavelength data available for the Abell 370 field is rich due to it being a Hubble Frontier Field (HFF) galaxy cluster \citep{Lotz_2017}. Although this cluster has extensive coverage across various instruments (including the {\sl Hubble Space Telescope} (\textit{HST}), \textit{Chandra}, VLA, and GMRT), this work focuses on the radio continuum and thermal X-ray components \citep{Feretti_2012}. We use MeerKAT UHF band observations for our primary analysis and spectral index derivation, comparing our findings to historical VLA and GMRT measurements.

\begin{figure*}
    \centering
    \includegraphics[width=\textwidth]{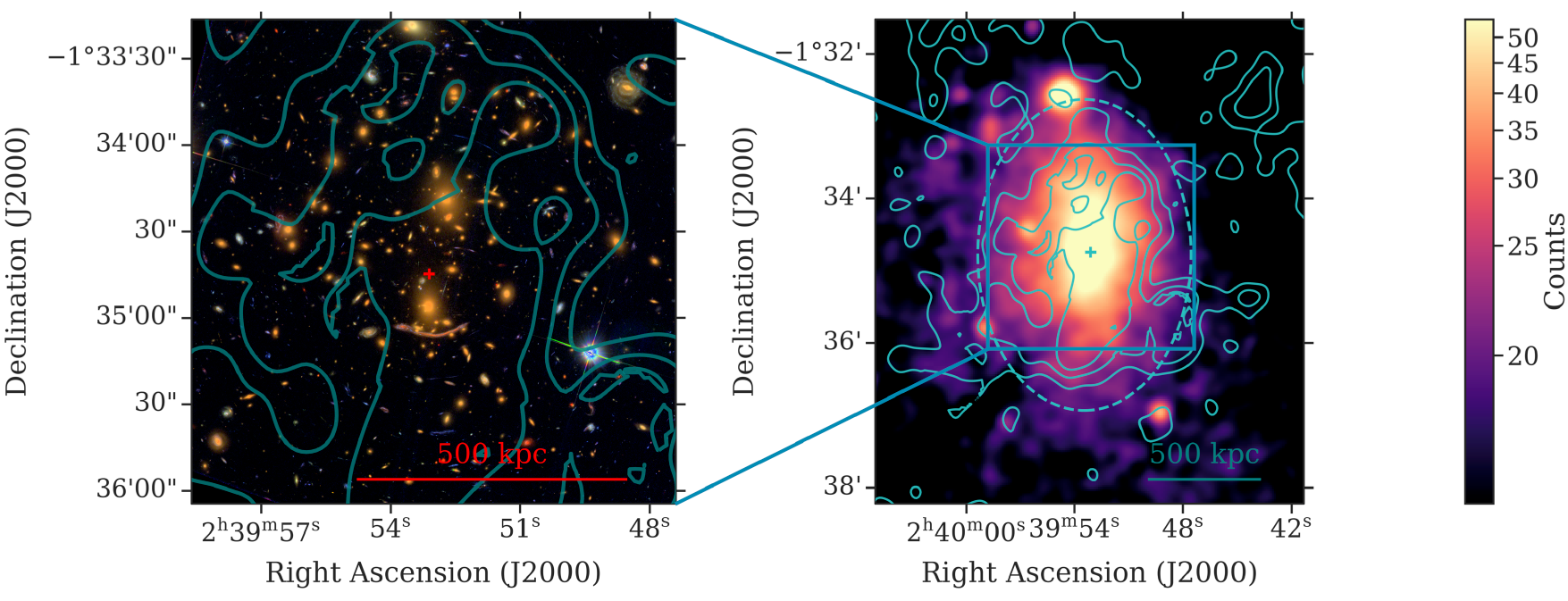}
    \caption{\textit{HST} \textit{RGB} image using the F814W, F606W, and F435W filters (left), and \textit{Chandra} X-ray map ($0.5\ \rm{to}\ 7.0\,\rm{keV}$, right) where the X-ray map is regridded to match the MeerKAT radio map and smoothed by a $12\,\rm{arcsec}$ Gaussian kernel. The X-ray map is colourised by counts, and we show the spectral index extraction region with a teal dashed ellipse. ACT SZ centroid is indicated with a cross, and radio contours from Fig.~\ref{fig:diffuse_emission} (right) are overlaid on both images. The 10 and 15$\sigma$ contours trace the X-ray emission closely, while the $5\sigma$ contour deviates in the southern part of the cluster.}
    \label{fig:ancil}
\end{figure*}

Fig.~\ref{fig:ancil} shows an \textit{HST} \textit{RGB} image using the F814W, F606W, and F435W filters, on the left with contours from the lower-resolution MeerKAT UHF-band map, as well as the \textit{Chandra} X-ray map with the same contours overlaid and pixels colourised by count, on the right. The \textit{Chandra} data comes from OBSID 515 (PI-Garmire) with the ACIS detector and was reduced with the Chandra Data Archive processing pipeline. The data was corrected for time-dependent gain, readout streaks, and background flares and has a final cleaned exposure time was 88 ks. The X-ray map was regridded to a $1.7\,\rm{arcsec}$ pixel scale and smoothed to match the resolution of the radio map. In both images, the ACT SZ centroid is indicated with a cross. The oval region in the right panel, shown by the teal dashed line, is constructed by approximately tracing the radio emission at $\sim5\sigma$. The integrated flux extracted from this region is $8.03 \pm 1.22\,\rm{mJy}$ ($816\,\rm{MHz}$ central frequency), where the uncertainties are calculated as the quadrature sum of a 15 per cent absolute flux calibration error and the statistical image noise ($\sigma_{\rm rms} \times \sqrt{N}$), with $N$ representing the number of synthesized beams within the extraction aperture.

The right panel in Fig.~\ref{fig:ancil} shows the diffuse radio emission closely matching the thermal X-ray emission within the teal dashed line region. This morphology provides evidence that Abell 370 conforms to the well-established radio-X-ray spatial correlation \citep{Govoni_2001b}. The X-ray and radio morphologies both exhibit a clear elongation in the north-south extent of the cluster. This spatial co-location, combined with the cluster's merging dynamical state, strongly suggests that the merger-induced turbulence is responsible for the re-acceleration of electrons resulting in synchrotron emission \citep{Brunetti_2001}. Notable X-ray point sources are visible in the field, with the most prominent being in the northeast.

The spatial co-location of diffuse radio and thermal X-ray emission provides evidence for the classification of the source as a radio halo. Additionally, this motivates the use of the constructed teal dashed region for flux extraction, to measure a spectral index using our MeerKAT UHF-band data.

\subsection{Spectral Index}
\label{spidx}

A key property of a radio halo is its spectral index, $\alpha$, which describes the power-law slope of its integrated flux density with frequency ($S_\nu\propto\nu^{\alpha}$). In this section, we use deep MeerKAT UHF-band observations to derive an in-band spectral index for the diffuse emission, and subsequently compare our measurement with values from the literature.

\begin{figure*}
    \centering
    \includegraphics[width=\textwidth]{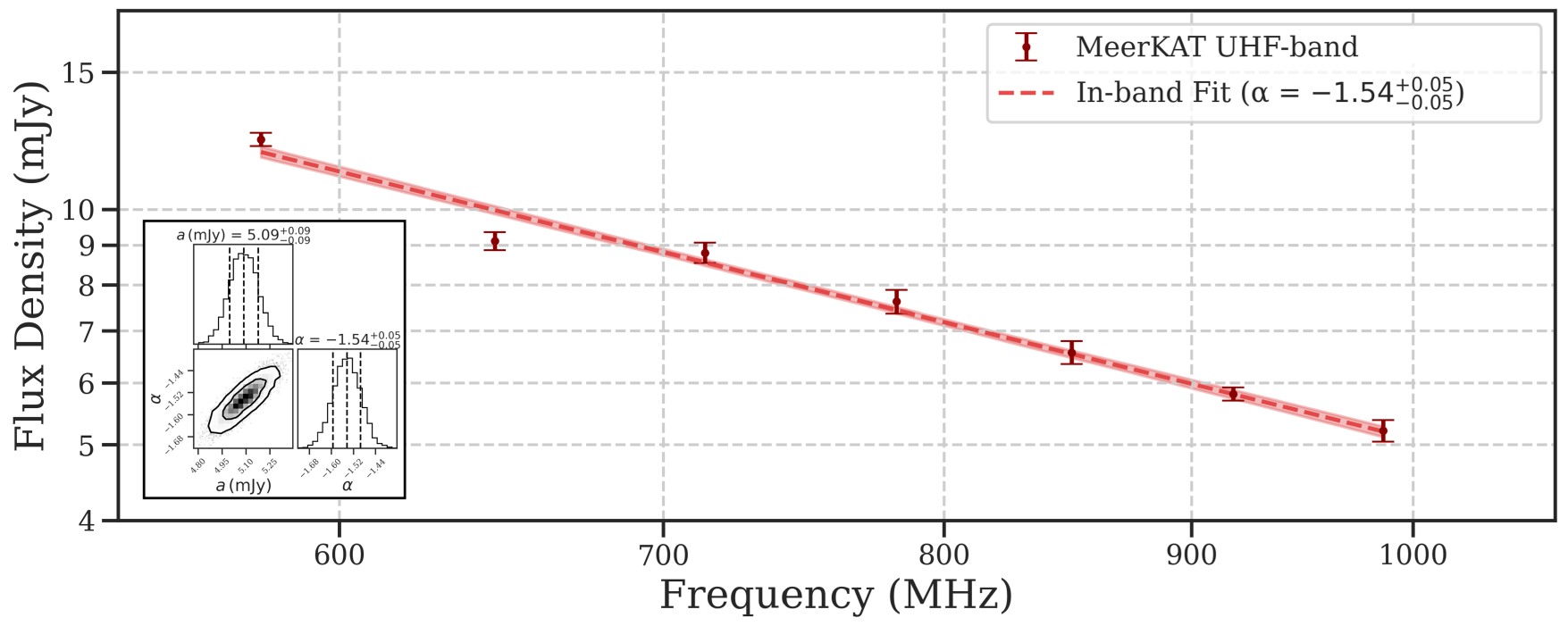}
    \caption{The integrated MeerKAT UHF in-band radio spectrum is shown in dark red. Flux density values are extracted within the specified oval region from seven in-band images, where uncertainties are derived as outlined in Sec.~\ref{spidx}. The red dashed line shows a power law model derived using MCMC \citep{Foreman_2013}, and returns a median posterior of $\alpha=-1.54^{+0.05}_{-0.05}$. MCMC confidence intervals ($1\sigma$) are indicated in shaded regions. In the bottom left, we show an inset of the 2D posterior probability distribution functions. The spectral index value indicates that the radio halo has a steeper spectrum than previously thought.}
    \label{fig:integrated_spec}
\end{figure*}

The integrated spectral index of the radio halo in Abell 370 is shown in Fig.~\ref{fig:integrated_spec}. Using \textsc{emcee} \citep{Foreman_2013} we determine spectral indices from the medians of posterior probability distributions from Markov Chain Monte Carlo (MCMC) runs with a model of $S_\nu = \rm{a}\nu^{\alpha}$, where $\rm{a}$ is the flux at $1\,\rm{GHz}$, $\rm{a}=S_{1\,\rm{GHz}}$. We run MCMC on in-band flux density from MeerKAT UHF-band images using the teal oval region in Fig.~\ref{fig:ancil} to extract flux values. Using seven sub-band images smoothed to a common PSF of $21\times21\,\rm{arcsec}$, we obtain a median of $\alpha=-1.54\pm0.05$. We indicate 68 per cent confidence intervals shaded in red and show the 2D and 1D posterior probability distribution of the spectral index. Uncertainties on flux densities are calculated as $\sigma=\sigma_{\rm{rms}}\times\sqrt{N}$, where $N$ is the number of beams in the teal region and $\sigma_{\rm{rms}}$ is determined from a source-free region in each sub-band image near the radio halo. These $\sigma_{\rm{rms}}$ estimates are consistent with the RMS measured within the teal region of the corresponding residual images. We list the sub-band fluxe densities in Table~\ref{tab:subband_fluxes}.

\begin{table}
\centering
\caption{Measured MeerKAT UHF in-band flux densities for the Abell 370 radio halo.}
\label{tab:subband_fluxes}
\begin{tabular}{ccc}
\hline
\hline
Sub-band & Frequency & Flux Density \\
 & (GHz) & (mJy) \\
\hline
1 & 0.578 & $12.30 \pm 0.24$ \\
2 & 0.646 & $9.12 \pm 0.24$ \\
3 & 0.714 & $8.81 \pm 0.27$ \\
4 & 0.782 & $7.63 \pm 0.26$ \\
5 & 0.850 & $6.57 \pm 0.22$ \\
6 & 0.918 & $5.81 \pm 0.12$ \\
7 & 0.986 & $5.22 \pm 0.17$ \\
\hline
\end{tabular}
\end{table}

The in-band MCMC probability distribution shows an ultra-steep median spectral index, consistent with $\alpha\le-1.5$ within uncertainties \citep{Cassano_2006,Brunetti_2008}. Given this steep spectrum, the non-detection at $3\,\rm{GHz}$ \citep{Xie_2020} is unsurprising due to the faint flux density predicted by the extrapolation. \citet{Xie_2020} reports $\alpha^{1500}_{325}=-1.10\pm0.09$ and an upper bound of $\alpha^{3000}_{1500}<-1.51$ using the VLA flux upper limit at $3\,\rm{GHz}$. These discrepancies could be due to our larger flux extraction region (motivated by a larger measured LLS) and/or the lower fidelity of the GMRT $325\,\rm{MHz}$ image.

Our discovery of the ultra-steep nature of the radio halo in Abell 370 significantly advances our understanding of this complex merging dynamical system and adds to the growing sample of USSRHs discovered to date \citep[e.g.][]{magolego_2026}, providing evidence in support of the turbulent re-acceleration model. We further explore and investigate this finding in the following sections, where we present the point-to-point radio-X-ray correlation analysis and spatial spectral index map for Abell 370. 

\section{Discussion}

\subsection{X-ray-radio Correlation}
\label{xray_discussion}

While the radio-X-ray spatial correspondence is qualitatively apparent in Fig.~\ref{fig:ancil}, a quantitative correlation can be established by performing a point-to-point analysis of the two maps. Typically, radio halos show positive correlation between radio and X-ray surface brightness, $I_{1.4\,\rm{GHz}}\propto\,I_{\rm{X}}^{k}$, where $I_{1.4\,\rm{GHz}}$ is the radio surface brightness at $1.4\,\rm{GHz}$ and $I_{\rm{X}}$ is the X-ray surface brightness, generally with a sub-linear slope, $k\le1$ \citep{Govoni_2001a,Feretti_2001,Giacintucci_2005,Vacca_2010,Hoang_2019,magolego_2026}. To explore this correlation, we regrid the \textit{Chandra} X-ray map to that of the radio image, then smooth it with a 12 arcsec Gaussian kernel.

\begin{figure*}
    \centering
    \includegraphics[width=\textwidth]{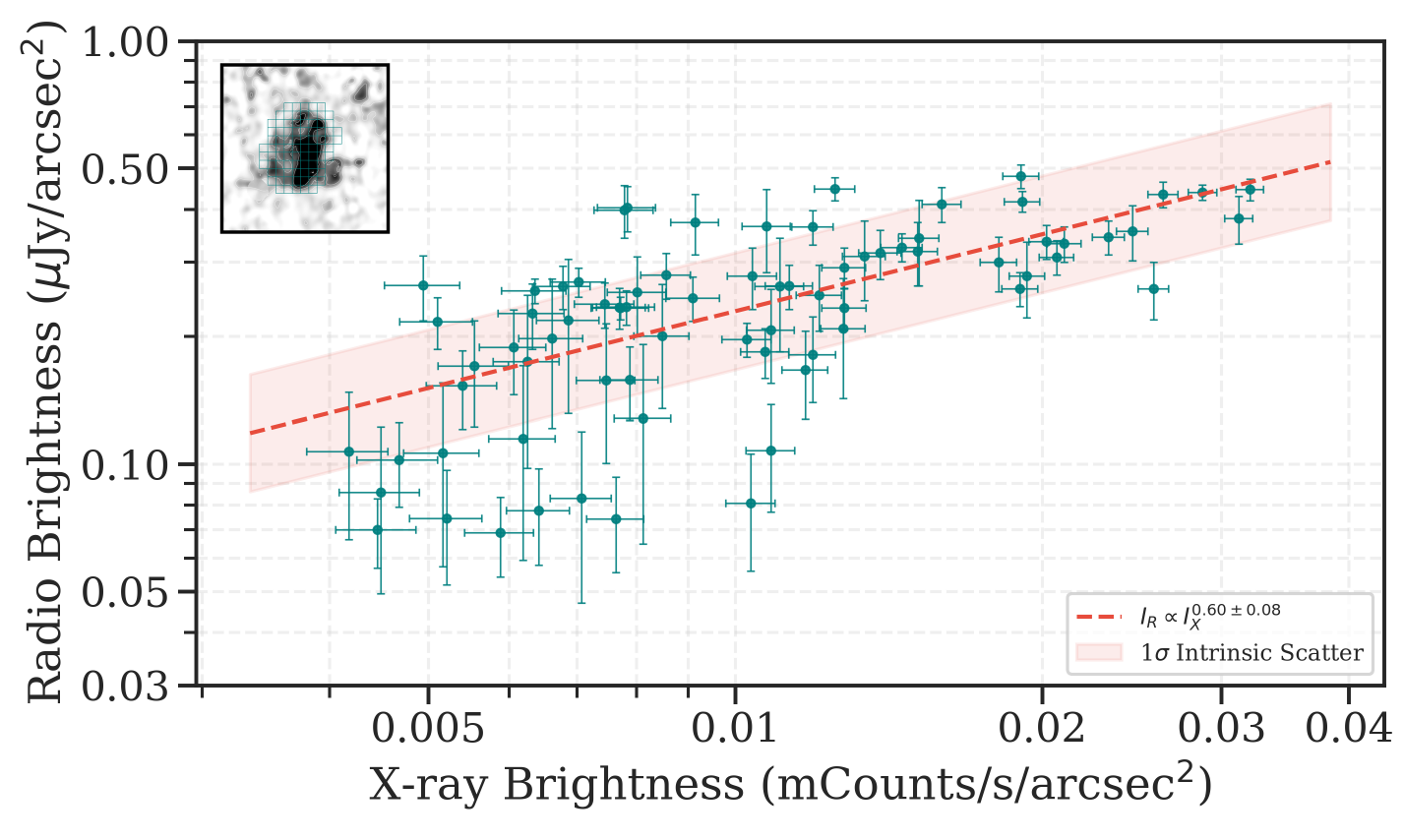}
    \caption{Radio to X-ray surface brightness correlation between MeerKAT UHF and \textit{Chandra} maps. Teal markers indicate the surface brightness measurements extracted across the cluster. The corresponding power-law fit, derived using Bayesian MCMC linear regression, is overlaid as a red dashed line. The lightly shaded red region represents the $1\,\sigma$ intrinsic scatter of the model. A radio inset is shown with the extraction regions overlaid in teal. The relation demonstrates a clear, statistically significant positive correlation.}
    \label{fig:xray}
\end{figure*}

We investigate the X-ray radio correlation in Fig.~\ref{fig:xray} by plotting radio surface brightness as a function of X-ray surface brightness for a set of regions covering the MeerKAT UHF-band radio halo. X-ray point sources are masked out and excluded from this analysis. Uncertainties on radio surface brightness markers are calculated as the standard deviation of the pixel values within each extraction region. We use a power law of the form $I_{\rm{radio}}\propto\,I_{\rm{X}}^{k}$ as our model. We filter the X-ray data to only include the soft energy band ($0.5-2\,\rm{keV}$) before proceeding with the analysis. X-ray surface brightness values are calculated as the sum of the pixel values normalised by the exposure time and physical area of the extraction region. Uncertainties on these values are derived using the square root of the variance (including the background counts) which is then normalised. The data points are further restricted to regions with a radio SNR $\ge 2.0$. We fit the aforementioned model to the data using Bayesian MCMC linear regression, finding $k=0.60\pm0.08$. The quoted uncertainties represent the $1\,\sigma$ standard deviation of the posterior, while the shaded regions indicate the $1\,\sigma$ intrinsic scatter.

The power-law slope of $k=0.60\pm0.08$ is consistent with literature region-based and pixel-by-pixel correlation analyses of radio halos \citep{Govoni_2001b,Feretti_2001,Giacintucci_2005,Brown_2011,Rajpurohit_2018,magolego_2026}. We find a Pearson correlation coefficient of $r=0.67$ and a Spearman correlation coefficient of $\rho=0.71$, both with $p\ll10^{-6}$, indicating a statistically significant, moderate-to-strong positive correlation. We have therefore qualitatively confirmed the visual correlation from Fig.~\ref{fig:ancil} with this region-based point-to-point correlation, confirming the physical connection between radio emission and thermal X-ray emission. However, the synchrotron emission itself is better explored with a spatial spectral index map.

\subsection{Spectral Index Map}
\label{spidx_spatial}

Previous studies of Abell 370 have not had sufficiently high SNR and image fidelity to create spectral index maps of the radio halo \citep{Xie_2020}. Here we present the spectral index map of the radio halo in Abell 370, demonstrating the high image fidelity of these MeerKAT UHF-band data.

\begin{figure}
    \centering
    \includegraphics[width=\columnwidth]{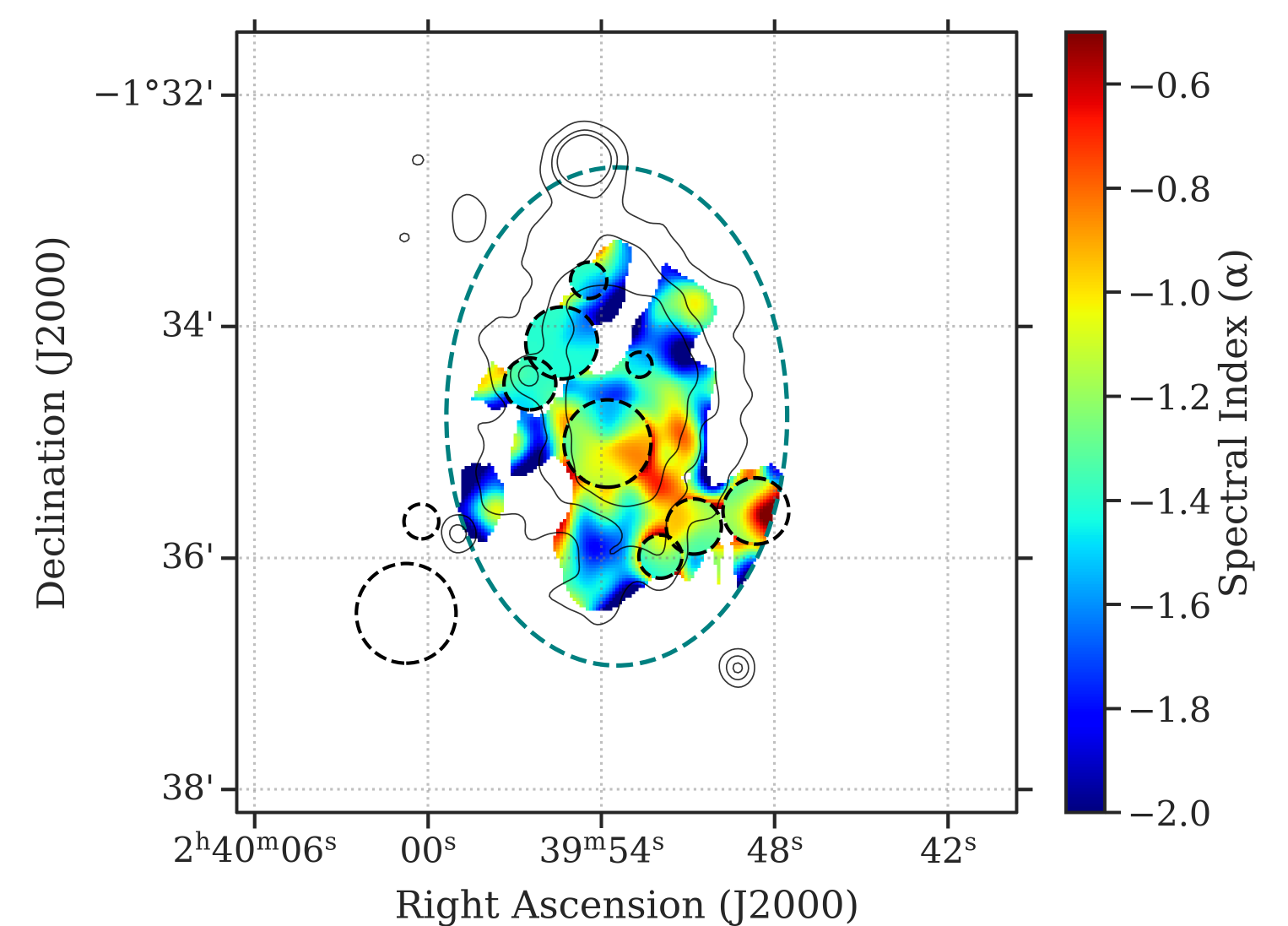}
    \caption{MeerKAT UHF in-band spectral index map of Abell 370 with X-ray contours ($1.5,\ 2,\ \rm{and} \ 3\times\sigma_{\rm{rms}}$) and a teal dashed line indicating the spectral index extraction region. Image plane subtracted radio point} sources are indicated with dashed black circles. We generate the map using \textsc{brats} \citep{Harwood_2013} with parameters outlined in Sec.~\ref{spidx_spatial} and seven in-band radio maps smoothed to a PSF of $21\times21\,\rm{arcsec}$. The pixels are colourised by the spectral index, saturated at $\alpha=-0.6$ and $\alpha=-2$. The majority of the radio halo shows a spectral index between -1 and -2, with a mean of -1.40.
    \label{fig:spidx_map}
\end{figure}

We generate the spectral index map using Broadband Radio Astronomy ToolS \citep[\textsc{brats}][]{Harwood_2013} in Fig.~\ref{fig:spidx_map}. We use seven sub-band images smoothed to a common PSF of $21\times21\,\rm{arcsec}$ and an SNR level in \textsc{brats} of \texttt{sigma=3} within the teal dashed region. The default "weighted least squares" fitting method is selected with the \texttt{specindexcalctype} command, together with a 15 per cent flux calibration error on all seven sub-band images. In Fig.~\ref{fig:spidx_map}, pixel values are colourised by spectral index, and contours are from the \textit{Chandra} X-ray map.

The \textsc{brats} spectral index map has a mean pixel value of $\alpha=-1.40$, which is broadly consistent with our integrated in-band spectral index of $\alpha=-1.54\pm0.05$ derived in Sec.~\ref{spidx} using the same smoothed sub-band images. While we lack sufficient SNR to fully populate the $5\sigma$ radio contours from Fig.~\ref{fig:diffuse_emission}, the core of the halo is reliably mapped. The map is relatively free of discrete point sources, demonstrating the success of our subtraction procedure. However, given how faint the diffuse radio emission is, deeper integration times are required to fully populate the spectral index map. A more complete map would enable a detailed, spatially resolved study of the radio--X-ray connection \citep{Botteon_2018}.

\subsection{Scaling Relations}
\label{scaling_relations}

The radio power-mass ($P_{1.4\,\rm{GHz}}-M_{500}$) scaling relation is a tight correlation, with low scatter, linking the mass of the cluster to the diffuse radio emission it hosts \citep{Cassano_2005}. While typical radio halos in galaxy clusters seldom deviate from this relation, USSRHs are known to be underluminous and lie below it \citep{Cassano_2013, Cuciti_2021b}. In this section, we contextualise the USSRH we find in Abell 370 by plotting its position on this relation. This allows us to compare it with known USSRHs and typical radio halos.

\begin{figure*}
    \centering
    \includegraphics[width=\textwidth]{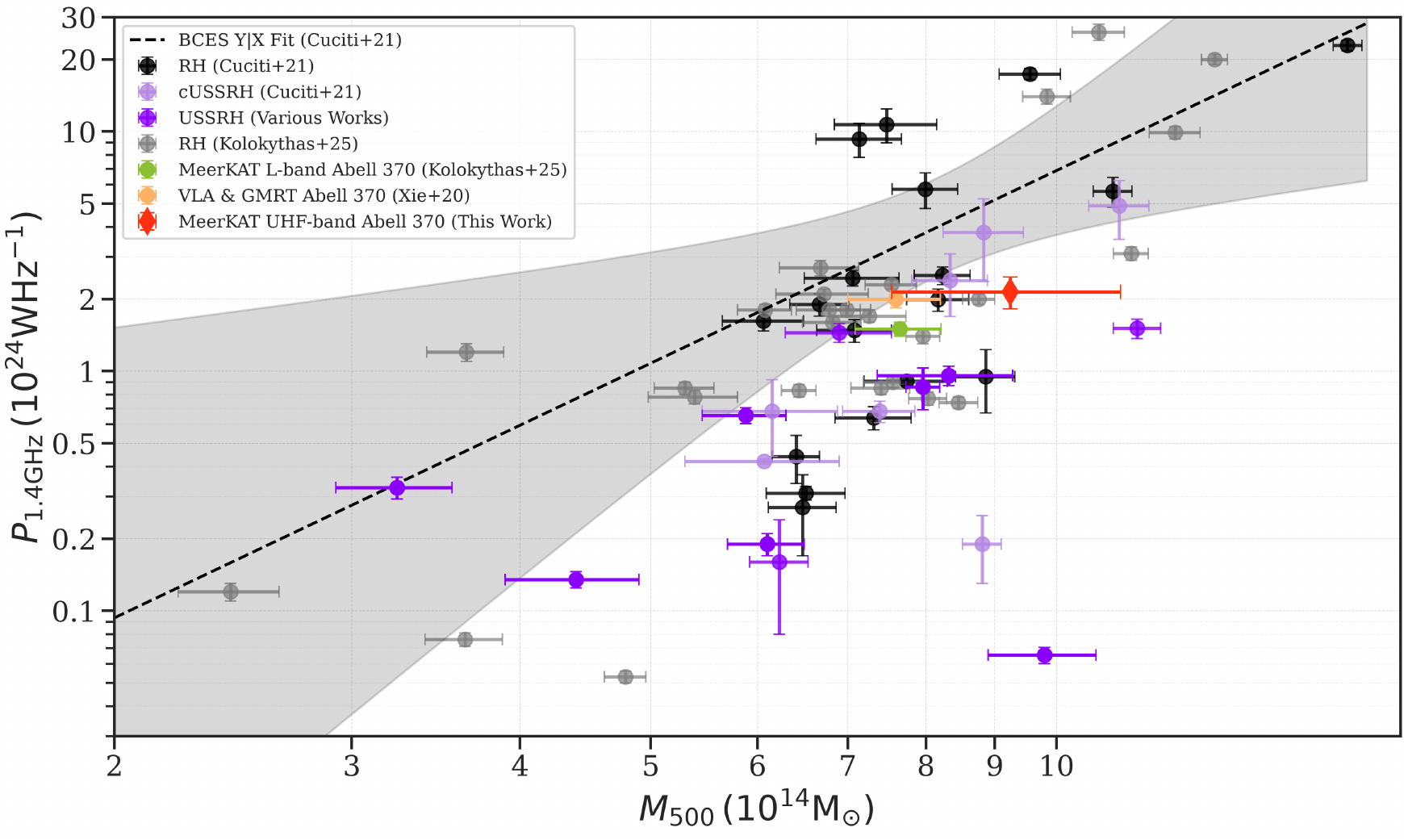}
    \caption{Radio power at $1.4\,\rm{GHz}$ ($P_{1.4\,\rm{GHz}}$) as a function of cluster mass ($M_{500}$) for radio halos and USSRHs. The red marker shows the MeerKAT UHF-band Abell 370 radio halo studied in this work, while the light green and orange markers indicate the MeerKAT L-band \citep{Kolokythas_2025} and VLA + GMRT \citep{Xie_2020} Abell 370 halo measurements, respectively. We denote the use of an ACT mass \citep{Hilton_2021} with a diamond marker, while Planck masses \citep{Planck_2016} are shown with circular markers. Dark and light purple markers indicate USSRHs and candidate USSRHs, respectively, compiled from various works \citep{Cuciti_2021,Duchesne_2021,Edler_2022,Pasini_2022,Venturi_2022,Rajpurohit_2023,magolego_2026}. Grey and black markers indicate radio halos from \citet{Kolokythas_2025} and \citet{Cuciti_2021}, respectively. The black dashed line represents the "RH only" BCES Y|X fit from the extended sample reported in \citet{Cuciti_2021b} (B=2.67 and A=0.07), with 95 per cent confidence intervals shown with shading. The radio halo in Abell 370 falls below this line and within the scatter of USSRHs and candidate USSRHs.}
    \label{fig:scaling_relation}
\end{figure*}

To standardize our measurement for literature comparison, we scale the observed UHF flux to a rest-frame frequency of $1.4\,\rm{GHz}$ using our derived spectral index of $\alpha=-1.54\pm0.05$. We then calculate a radio power using:
\begin{equation}
    P_{1.4\,\rm{GHz}} = 4\pi D^2_L S_{1.4\,\rm{GHz}} (1+z)^{-(\alpha+1)},
    \label{eq:power}
\end{equation}
where $D_L$ is the luminosity distance to the halo, and the rest are as previously defined. This yields $P_{1.4\,\rm{GHz}} = (2.14 \pm 0.33) \times 10^{24}\,\rm{W\,Hz^{-1}}$. The position of Abell 370 in the context of the radio power at $1.4\,\rm{GHz}$ ($P_{1.4\,\rm{GHz}}$) - mass ($M_{500}$) scaling relation is investigated in Fig.~\ref{fig:scaling_relation}. Abell 370 measurements are plotted for this work in red, from \citet{Kolokythas_2025} in light green, and from \citet{Xie_2020} in orange. We use the weak lensing corrected mass from the ACT catalogue \citep{Hilton_2021}, $M_{500}(10^{14}\rm{M_\odot})= 9.24^{+1.92}_{-1.69}$, while the latter literature measurements use the mass from \citet{Planck_2016}, $M_{500}(10^{14}\rm{M_\odot})= 7.6\pm0.6$. We differentiate the masses with marker shape: diamond for ACT and circular for Planck. Note that the uncorrected mass, $M_{500}(10^{14}\rm{M_\odot})= 7.4^{+1.4}_{-1.2}$, from ACT is more consistent with the mass from \citet{Planck_2016}, however, motivated by the well constrained mass modeling of Abell 370. We plot a literature compilation of USSRHs and candidate USSRHs (cUSSRH) in dark and light purple, respectively \citep{Cuciti_2021,Edler_2022,Duchesne_2021,magolego_2026,Rajpurohit_2023,Venturi_2022,Pasini_2022}, as well as a non-exhaustive sample of radio halos in black and grey \citep{Kolokythas_2025,Cuciti_2021}. Over-plotted, we show the BCES Y|X fit by \citet{Cuciti_2021b} for radio halos only in their extended sample, with coefficients of B=2.67 and A=0.07. The 95 per cent confidence interval is calculated as outlined in \citet{Cuciti_2021} and shown with grey shading. For literature sources where $1.4\,\rm{GHz}$ measurements are unavailable, we first extrapolate the nearest quoted flux density to $1.4\,\rm{GHz}$ using the standard relation $S \propto \nu^{\alpha}$. We then calculate the rest-frame radio power using equation~\ref{eq:power}.

Fig.~\ref{fig:scaling_relation} shows that USSRHs and cUSSRHs are preferentially found below the scaling relation fit shown here (extended sample and RHs only), and indeed below other fits presented in \citet{Cuciti_2021b}. All but two cUSSRHs \citep{Cuciti_2021} and one USSRH \citep{Edler_2022} lie below the 95 per cent confidence region of this relation. Our own measurement, and two literature measurements \citep{Xie_2020,Kolokythas_2025}, all place Abell 370 below this fit and outside the 95 per cent confidence region if uncertainties are taken into account. If a Planck-derived mass is adopted, our measurement of Abell 370 aligns closely with the findings of \citet{Xie_2020}. While the halo is then within uncertainties of the 95 per cent confidence interval of the BCES Y$\vert{}$X fit \citep{Cuciti_2021b}, it remains relatively underluminous. Crucially, these results stem from three independent flux measurements and two independent mass measurements. 

We confirm that in the radio power-mass scaling relation, USSRHs occupy a different parameter space than typical RHs. The fact that Abell 370 is underluminous and lies well within the scatter of this USSRH population provides strong supporting evidence for its ultra-steep spectrum nature discussed in Sec.~\ref{fig:integrated_spec}.

\subsection{Classification}
\label{classification}

With evidence from the integrated spectral index (Fig.~\ref{fig:integrated_spec}), the \textsc{brats} spectral index map (Fig.~\ref{fig:spidx_map}), and the location of Abell 370 on the $P_{1.4\,\rm{GHz}}-M_{500}$ scaling relation (Fig.~\ref{fig:scaling_relation}) we can now confidently classify the radio halo in Abell 370 as an USSRH. Before presenting this final classification, we perform a qualitative cross-check between MeerKAT observations to validate our in-band spectral indices and absolute flux calibration.

\begin{figure}
    \centering
    \includegraphics[width=\columnwidth]{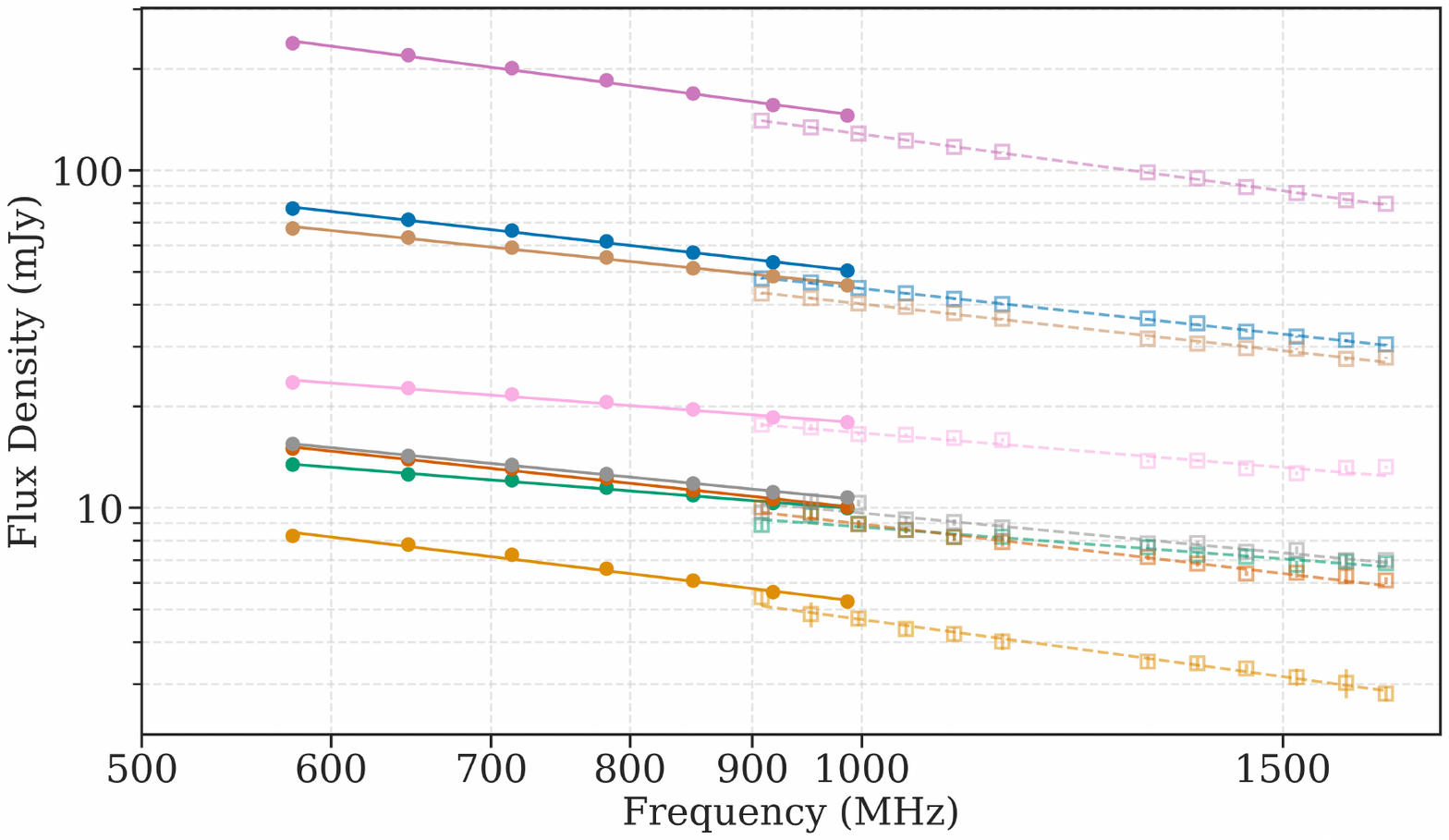}
    \caption{MeerKAT UHF and L-band spectra of six isolated radio point sources in the Abell 370 field, with each colour representing a source. We use MCMC \citep{Foreman_2013} to fit a power law and derive a spectral index using seven smoothed in-band UHF-band images and 12 in-band L-band images separately. A fractional bias is apparent between the L- and UHF-band absolute calibration, however, this lies within the observatory-derived uncertainties.}
    \label{fig:spidx_consistency}
\end{figure}

The purpose of this exercise is to qualitatively demonstrate spectral index consistency across MeerKAT UHF-band and L-band. Fig.~\ref{fig:spidx_consistency} plots the integrated spectral index of eight radio point sources that are confirmed manually to be free from source confusion in the Abell 370 field of view. We use seven in-band MeerKAT UHF-band images as well as the 12 sub-band L-band images from MGCLS archival data\footnote{\url{https://archive-gw-1.kat.ac.za/public/repository/10.48479/7epd-w356/index.html}} \citep{Knowles_2022} and perform the same MCMC sampling on these data. The resulting mean spectral index for these sources, differentiated by colour, is $\alpha=-0.74\pm0.13$ for the UHF-band and $\alpha=-0.76\pm0.15$ for the L-band. This agreement was achieved after correcting for an initial systematic offset by applying the updated flux calibrator model from the MeerKAT documentation\footnote{\url{https://skaafrica.atlassian.net/wiki/spaces/ESDKB/pages/1481408634/Flux+and+bandpass+calibration}}.

As shown in Fig.~\ref{fig:spidx_consistency}, the derived spectral indices exhibit consistency ($\Delta\alpha=0.02$) and successfully recover the canonical point-source spectral index of $\alpha\sim-0.7$ \citep{Condon_1992}. This, coupled with the low scatter in the derived spectral indices, reinforces the reliability of both the UHF- and L-band MeerKAT measurements. While we note a slight amplitude offset, likely arising from differences in visibility weighting and tapering, this merely shifts the normalisation without affecting the measured slopes or the scientific conclusions of this work.

Taken together, our analysis provides multiple lines of evidence to classify the diffuse emission in Abell 370. First, its LLS of $1.1\,\rm{Mpc}$ and strong spatial correlation with the thermal X-ray emission (see Sec.~\ref{diff_radio} and Sec.~\ref{xray_discussion}) confirm its nature as a radio halo \citep{Govoni_2001b}. We further refine this classification based on our in-band spectral index ($\alpha=-1.54\pm0.05$) and the source's position within the USSRH population on the $P_{1.4\,\rm{GHz}}$-$M_{500}$ scaling relation (Fig.~\ref{scaling_relations}), as we classify this source as an USSRH. This refines its previous classifications from a radio halo \citep{Xie_2020} or a candidate radio halo \citep{Knowles_2022} to a confirmed USSRH. 

\section{Conclusions}
\label{conclusions}

In this work, we use MeerKAT UHF-band data, from a project aimed at high-$z$ lensed \textsc{Hi} emission, to deliver commensal diffuse radio halo science. We uncover a significantly larger spatial extent of diffuse radio emission in Abell 370 with deep MeerKAT observations and meticulous point source subtraction. The derived spectral index is consistent with an USSRH. We find a strong qualitative and quantitative radio-X-ray correlation within the halo, and confirm the ultra-steep spectrum with a spectral index map and the power-mass scaling relation. Our main findings are as follows:

\begin{enumerate}
    \item Our tapered imaging reveals a radio halo with an LLS of $\sim1.1\,\rm{Mpc}$, larger than previously measured. Additionally, we find radio extensions to the southeast and southwest.
    \item We find an in-band integrated spectral index of $\alpha=-1.54\pm0.05$, consistent with an USSRH.
    \item The dynamically unrelaxed radio halo exhibits a strong radio-X-ray correlation of $k=0.60\pm0.08$, supporting the turbulent re-acceleration model.
    \item We create a spectral index map of the radio halo in Abell 370, confirming its ultra-steep-spectrum.
    \item In the context of the power-mass scaling relation for radio halos, Abell 370 is positioned within the scatter of known USSRHs, providing further evidence for its classification.
\end{enumerate}

This work showcases the compelling commensal radio continuum science possible with deep MeerKAT programs focused on targeting massive clusters in order to detect gravitationally lensed \textsc{Hi} sources in the distant universe.

\section*{Acknowledgements}

We thank the anonymous referee for a careful review and insightful feedback that helped improve this work. We would like to thank the staff of the South African Radio Astronomy Observatory (SARAO\footnote{www.sarao.ac.za}) who made these observations possible. We thank Isaac Magolego and Dakalo Phuravhathu for insightful conversations and assistance in data reduction and point source subtraction. GDL and RPD acknowledge funding from the South African Radio Astronomy Observatory (SARAO), which is a facility of the National Research Foundation (NRF), an agency of the Department of Science, Technology and Innovation (DSTI). RPD acknowledges funding by the South African Research Chairs Initiative of the DSI/NRF (Grant ID: 77948). The MeerKAT telescope is operated by the South African Radio Astronomy Observatory, which is a facility of the National Research Foundation, an agency of the Department of Science and Innovation. We acknowledge the use of the \textsc{ILIFU}\footnote{www.ilifu.ac.za} cloud computing facility - www.ilifu.ac.za, a partnership between the University of Cape Town, the University of the Western Cape, Stellenbosch University, Sol Plaatje University, the Cape Peninsula University of Technology and the South African Radio Astronomy Observatory. The \textsc{ILIFU} facility is supported by contributions from the Inter-University Institute for Data Intensive Astronomy (IDIA - a partnership between the University of Cape Town, the University of Pretoria and the University of the Western Cape), the Computational Biology division at UCT and the Data Intensive Research Initiative of South Africa (DIRISA). This work has made use of the Cube Analysis and Rendering Tool for Astronomy \citep[CARTA,][]{Comrie_2021}. This research made use of Astropy\footnote{http://www.astropy.org}, a community-developed core Python package for Astronomy. This research makes use of the semi-automated MeerKAT data calibration pipeline, \textsc{oxkat}\footnote{https://github.com/IanHeywood/oxkat}.

\section*{Data Availability}

The raw visibilities used in this work are from project ID SCI-20241101-GL-01 on the MeerKAT archive\footnote{https://archive.sarao.ac.za}. The authors may make raw visibilities and data products available upon reasonable request. 



\bibliographystyle{mnras}
\bibliography{example} 








\bsp	
\label{lastpage}
\end{document}